\documentclass{PoS}

\title{Resolving the Innermost Geometry of Relativistic Jets in Active Galactic Nuclei}

\ShortTitle{Resolving the Geometry of the Innermost Relativistic Jets in AGNi}

\author {\speaker{Juan~Carlos Algaba}$^{a,b}$, {Masanori Nakamura}$^{c}$, {Keiichi Asada}$^{c}$, {Sang~Sung Lee}$^{b,d}$\\
\llap{$^a$}Department of Physics and Astronomy, Seoul National University, 1 Gwanak-ro, Gwanak-gu, Seoul 08826, Republic of Korea\\
\llap{$^b$}Korea Astronomy \& Space Science Institute, 776, Daedeokdae-ro, Yuseong-gu, Daejeon, 305-348, Republic of Korea\\
\llap{$^c$}Academia Sinica, Institute of Astronomy and Astrophysics, 11F of Astronomy-Mathematics Building, AS/NTU No. 1, Taipei 10617, Taiwan\\
\llap{$^d$}Korea University of Science and Technology, 217 Gajeong-ro, Yuseong-gu, Daejeon 34113, Republic of Korea\\
E-mail: \email{algaba@astro.snu.ac.kr}}

\abstract{In the current paradigm, it is believed that the compact VLBI radio core of radio-loud active galactic nuclei (AGNi) represents the innermost upstream regions of relativistic outflows. These regions of AGN jets have generally been modeled by a conical outflow with a roughly constant opening angle and flow speed. Nonetheless, some works suggest that a parabolic geometry would be more appropriate to fit the high energy spectral distribution properties and it has been recently found that, at least in some nearby radio galaxies, the geometry of the innermost regions of the jet is parabolic. We compile here multi-frequency core sizes of archival data to investigate the typically unresolved upstream regions of the jet geometry of a sample of 56 radio-loud AGNi. Data combined from the sources considered here are not consistent with the classic picture of a conical jet starting in the vicinity of the super-massive black hole (SMBH), and may exclude a pure parabolic outflow solution, but rather suggest an intermediate solution with quasi-parabolic streams, which are frequently seen in numerical simulations. Inspection of the large opening angles near the SMBH and the range of the Lorentz factors derived from our results support our analyses. Our result suggests that the conical jet paradigm in AGNi needs to be re-examined by millimeter/sub-millimeter VLBI observations.}

\FullConference{14th European VLBI Network Symposium \& Users Meeting (EVN 2018)\\
		8-11 October 2018\\
		Granada, Spain}

\begin{document}

\section{Introduction}
In the current paradigm, it is believed that the compact VLBI radio core of radio-loud AGNi represents the innermost upstream regions of relativistic outflows. These regions of AGN jets have generally been modeled by a conical outflow with a roughly constant opening angle and flow speed. Nonetheless, some works \cite{AsadaNakamura12,Tseng16,Akiyama18,Nakahara18a,Nakahara18b} suggest that a parabolic geometry would be more appropriate. It is thus reasonable to consider if the classic paradigm has to be revisited.

Unfortunately, in most of the sources, the transverse section of the jet is not well resolved and it is difficult to check if this trend is a paradigm for all AGNi. In order to study the jet sizes upstream, we take advantage of the core shift. On one hand, The core can be used as a probe for the properties of the upstream jet regions. On the other hand, the core shift provides the location of this region with respect to the central engine. Thus, by probing the core size, we can actually investigate the upstream jet size at distance given by the core shift. The jet geometry can then be parametrized with transverse radius $R\propto r^{\epsilon}$  (with $r$, distance from the nucleus). The jet half opening angle $\theta=\arctan(R/r)$ is thus constant and is representative of a conical; ballistic jet if $\epsilon=1$, or decreases in a parabolic; collimating jet if $0<\epsilon<1$. 

We investigated archival data of various surveys (see \cite{Algaba16} for details) in order to obtain information of VLBI core sizes at 1.6, 2.3, 5.0, 8.6, 15, 22 and 86~GHz. We also estimated the location of the core from the central engine at each frequency for every source using the core shift values in \cite{Pushkarev12}. We were able to compile appropriate values for the core size and distance for at least four frequencies in a total of 56 objects.

\section{Results}
\begin{figure}[ht]
\center
\includegraphics[scale=0.40,trim={0.3cm 0cm 0cm 0cm},clip]{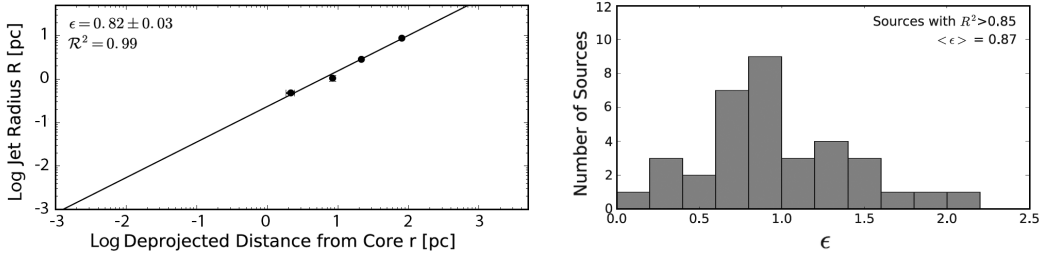}
\caption{\textbf{Left:} Fit for the core transverse size versus core shift distance for 2128--123. \textbf{Right: } Histogram of the fitted values of $\epsilon$ for the sources with $R^2\geq0.85$.} 
\label{fig1}
\end{figure}

In Figure \ref{fig1} left, we show a sample plot of the transverse size versus the core shift distance for 2128--123. A considerable number of sources show a large scatter in their data, with the goodness of the fit $R < 0.85$. We consider the possibility that these  low values may be due to the lack of significant data for proper statistics and a reliable fit. In order to construct a criteria to check for the data scatter, we consider these fits with
$R^2\geq0.85$. Under this criteria, 11\%, 60\%, and 29\% of the sources have small ($\epsilon < 1/2$), intermediate ($1/2 < \epsilon < 1$), and
large ($\epsilon > 1$) geometry values, respectively. 

If we restrict ourselves to only these sources with $R^2\geq0.85$, then the average value for the geometry parameter is  $<\epsilon> = 0.85$. In Figure \ref{fig1} right, we show a histogram for the values of $\epsilon$ for such sources. A Kolmogorov-Smirnov test indicates that the distribution is different from a Gaussian with a significance of 90\%. Thus, although with caveats and limited amount of data, our source-by-source results suggest that there is a significant peak in the number of sources following a semi-parabolic geometry.

\section{A Global View}
As a way to avoid the constraints due to the limited amount of data for each source, we consider to combine the data for all sources together. Although our sample may consist on a mixture of different geometries and a fit may not be relevant, this will allow us to study the global behavior of all sources and to search for a common structural trend. In order to perform a proper comparison, data points are converted into the units of the gravitational radius $r_g = GM_{BH}/c^2$ by using the black hole mass $M_{BH}$ tabulated in \cite{WooUrry02} and \cite{Zamaninasab14}.

\begin{figure}[ht]
\center
\includegraphics[scale=0.85,trim={0.3cm 0cm 0cm 0cm},clip]{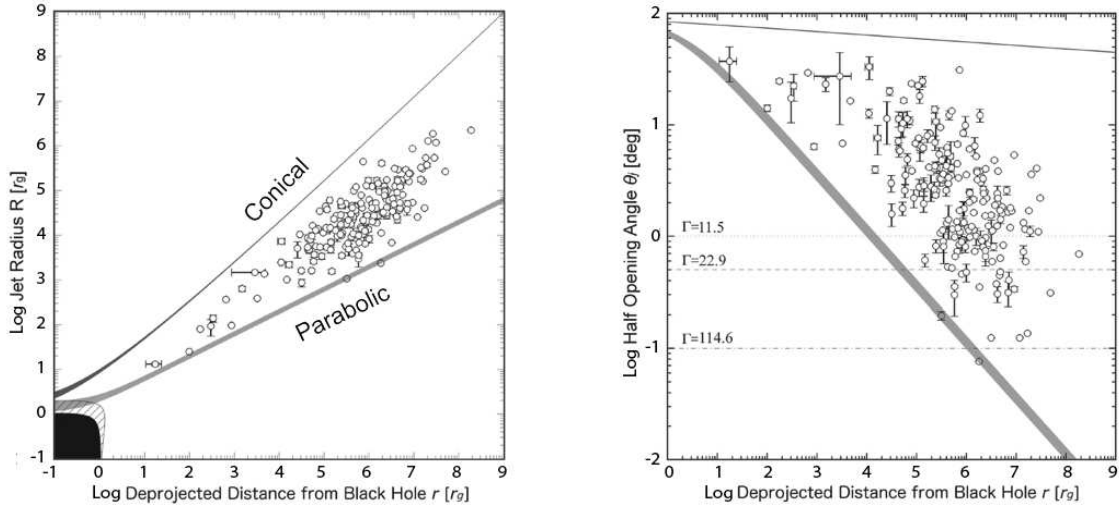}
\caption{The light gray area denotes the genuine parabolic streamline, while the dark gray area denotes the quasi-conical streamline. A variation from $0.5<a<0.998$ is considered as a shaded area. Left: Filled black region denotes the black hole, while the hatched area represents the ergosphere for the black hole spin parameter a=0.998. Right: Dotted, dashed, and dotted?dashed lines show Lorentz factors for 1$^{\circ}$, 0$^{\circ}$. 5, and 0$^{\circ}$.1, respectively.} 
\label{fig2}
\end{figure}

The collected data (see Figure \ref{fig2} left) may not support the classical picture of a jet starting from the vicinity of the black hole with neither (1) a conical geometry ($\epsilon = 1$) nor (2) a genuine parabolic geometry ($\epsilon = 1/2$). Instead, the data fits in an intermediate region where semi-parabolic streamlines ($0.5 < \epsilon < 1$) would exist. Quasi-parabolic streamline can be generally formed in radiatively efficient accretion flows by utilizing general relativistic radiation magnetohydrodynamics (GRRMHD) simulations (e.g., \cite{McKinney15, SadowskiNarayan15,Nakamura18}).

We derived the jet intrinsic half opening angles. In Figure \ref{fig2}, right we plot these as a function of distance from the central engine. It is clear that, for small radii, the opening angles are quite large, suggesting that a quasi-conical expansion is unlikely in such a regime (otherwise jets
would be unrealistically wide even near the jet base). This also supports our view in which quasi-parabolic structures are common.

The jet is expected to be causally connected with its symmetry axis, implying $\Gamma\theta<1$, with $\Gamma$ the Lorentz Factor \cite{Komissarov09,Tchekhovskoy09}. Assuming $\Gamma\theta\sim0.2$ \cite{ClausenBrown13}, the derived values of $\Gamma\sim10-20$ that we find are in agreement with values of $\Gamma\sim15$ from observations \cite{Jorstad05, Hovatta09}, which indicates that the derivation of half opening angles from core size analyses is reasonable.

\section{Conclusions}
With our criteria, 60\% of the sources show quasi-parabolic structure, with $1/2 < \epsilon < 1$, and the median geometry value is $<\epsilon> = 0.85$. Furthermore, the combined data fits in a region between genuine parabolic and conical geometries, supporting the idea that, near the vicinity of the central engine, the jet exhibits a semi-parabolic geometry. This seems to suggest that a semi-parabolic jet shape may be more common near the innermost few parsecs of the jet, in contrast with the conical shapes typically found on deca-parsec scales or further.  We speculate that a transition from parabolic to conical geometry may occur. Our result suggests that the conical jet paradigm in AGNi needs to be re-examined by millimeter/sub-millimeter VLBI observations.

\end{document}